\documentclass[times,twocolumn,final]{elsarticle}

\usepackage{medima}
\usepackage{framed,multirow}

\usepackage{latexsym}

\usepackage{amsmath,amssymb,amsfonts}
\usepackage{algorithmic}
\usepackage{graphicx}
\usepackage{textcomp}

\usepackage{microtype}
\usepackage{multirow}
\usepackage{booktabs}
\usepackage{rotating}
\usepackage{soul}
\usepackage{cancel}

\usepackage{url}
\usepackage{xcolor}

\usepackage{hyperref}

\definecolor{color}{rgb}{.8,.349,.1}

\journal{Medical Image Analysis}

\begin{document}

\verso{S. Yang, X. Wu, S. Ge \textit{et~al.}}

\begin{frontmatter}

\title{Knowledge Matters: Chest Radiology Report Generation with General and Specific Knowledge}

\author[1,4]{Shuxin \snm{Yang}\fnref{fn1}}

\author[3]{Xian \snm{Wu}\corref{cor1}}
\author[3]{Shen \snm{Ge}}
\author[1,2]{S. Kevin \snm{Zhou}\corref{cor1}}
\author[1,4]{Li \snm{Xiao}\corref{cor1}}

\cortext[cor1]{Corresponding author: 
  andrew.lxiao@gmail.com,xiaoli@ict.ac.cn (X. Li); skevinzhou@ustc.edu.cn (S. K. Zhou); kevinxwu@tencent.com(X. Wu).}
  
\address[1]{Key Lab of Intelligent Information Processing of Chinese Academy of Sciences (CAS)
Institute of Computing Technology, CAS, Beijing, 100190, China}
\address[2]{School of Biomedical Engineering \& Suzhou Institute for Advanced Research
Center for Medical Imaging, Robotics, and Analytic Computing \& LEarning (MIRACLE) 
University of Science and Technology of China, Suzhou 215123, China
}
\address[3]{Tencent Medical AI Lab, Beijing, 100094, China}
\address[4]{University of Chinese Academy of Sciences, Beijing, 100049, China}

\fntext[fn1]{This work was done when Shuxin Yang was an intern at Tencent Medical AI Lab.}


\begin{abstract}
Automatic chest radiology report generation is critical in clinics which can relieve experienced radiologists from the heavy workload and remind inexperienced radiologists of misdiagnosis or missed diagnose. Existing approaches mainly formulate chest radiology report generation as an image captioning task and adopt the encoder-decoder framework. However, in the medical domain, such pure data-driven approaches suffer from the following problems: 1) visual and textual bias problem; 2) lack of expert knowledge. In this paper, we propose a knowledge-enhanced radiology report generation approach introduces two types of medical knowledge: 1) General knowledge, which is input independent and provides the broad knowledge for report generation; 2) Specific knowledge, which is input dependent and provides the fine-grained knowledge for chest x-ray report generation. To fully utilize both the general and specific knowledge, we also propose a knowledge-enhanced multi-head attention mechanism. By merging the visual features of the radiology image with general knowledge and specific knowledge, the proposed model can improve the quality of generated reports. The experimental results on the publicly available IU-Xray dataset show that the proposed knowledge-enhanced approach outperforms state-of-the-art methods in almost all metrics. And the results of MIMIC-CXR dataset show that the proposed knowledge-enhanced approach is on par with state-of-the-art methods. Ablation studies also demonstrate that both general and specific knowledge can help to improve the performance of chest radiology report generation.

\end{abstract}


\end{frontmatter}



\section{Introduction}
\label{sec:introduction}

Chest radiology report generation targets to generate a paragraph to address the observations and findings of a given radiology image, writing a radiology report is time-consuming and error-prone. It takes on average 10 min or more based on the radiologist's degree of experience~\citep{alfarghaly2021automated}, and even skilled and experienced radiologists fail to note important findings on 30\% of chest radiographs that are positive for disease and also have a false-positive rate of approximately 2\% for negative cases~\citep{bruno2015understanding}. Therefore automatically generating radiology reports is of great clinical value. On the one hand, it can relieve experienced radiologists from the heavy workload. The radiologists can modify the generated draft rather than start from scratch; On the other hand, it can remind the inexperienced radiologists of the potential abnormalities, thus avoiding misdiagnosis and missed diagnosis. 

Due to its clinical importance, automatic chest radiology report generation has attracted growing research interests~\citep{zhou2021review}. Existing works~\citep{xu2015show, lu2017knowing, anderson2018bottom, Jing2018On,Yuan2019Automatic} mainly formulate radiology report generation as an image captioning task and adopt the encoder-decoder framework: firstly extract visual features from the radiology image with the encoder and then generate the textual report with the decoder. 
The results of previous works~\citep{xu2015show, lu2017knowing, anderson2018bottom, Jing2018On, Yuan2019Automatic} show that an encoder-decoder framework has achieved decent performance gain on radiology report generation. For example, the proposed encoder-decoder model of \cite{Jing2018On} achieves 0.154 of BLEU-4 score. However, it still has two problems: 1) visual and textual bias: the model is expected to pay more attention to the abnormalities when generating the radiology report. However, in most cases, the abnormal regions only occupy a small part of the radiology image. The descriptions relating to these abnormal regions only occupy a small fraction in the final report. As a result, the pure data-driven encoder-decoder approaches could lean towards normal descriptions and fail to capture the abnormalities; 2) Lack of expert knowledge: the pure encoder-decoder based approaches could not incorporate the expert knowledge precisely for generating reports, which makes it hard to be a widely adopted tool in clinics. 

To address the above problems, several approaches have been proposed to integrate medical knowledge in modelling. For example, Hybrid Retrieval-Generation Reinforced(HRGR)~\citep{Li2018Hybrid} builds a template database based on prior knowledge by manually filtering a set of sentences in the training corpus; Meets Knowledge Graph(MKG)~\citep{Zhang2020when} and Posterior and Prior Knowledge Exploring and Distilling(PPKED)~\citep{Liu2021Exploring} incorporate a manual pre-constructed knowledge graph to enhance the generation. However, the scope of knowledge introduced by these works is quite limited. For example, the pre-constructed knowledge graph~\citep{Liu2021Exploring} only includes 20 entities that are insufficient for radiology report generation. On the other hand,  the RadGraph, which contains 6 million entities, is a large-scale knowledge graph extracted from the MIMIC-CXR dataset, which can better assist radiology report generation.

To better integrate existing knowledge into chest radiology report generation, we categorize medical knowledge into general knowledge and specific knowledge based on the RadGraph. In the following, we define the general knowledge as image-independent knowledge from a standard knowledge base; We also define the specific knowledge as image-dependent knowledge that relates to the current input image: for each radiology image, we retrieve similar images and collect a customized set of knowledge from their reports. To merge the general knowledge, specific knowledge, and the visual features from the radiology image, we propose a knowledge-enhanced multi-head attention mechanism to model the structural knowledge information. 

We evaluate the proposed method on two public datasets: IU-Xray and MIMIC-CXR. Besides the typical natural language generation (NLG) metrics, we adopt clinical efficacy (CE) metrics to analyze the quality of generated reports in clinics. The results show that the proposed method achieves state-of-the-art performance on both NLG and CE metrics. The results also indicate that the radiology report generation benefits from both general knowledge and specific knowledge. 

The main contributions are as follows:
\begin{itemize}
    \item We propose a novel radiology generation framework with two extra knowledge: general knowledge and specific knowledge. The general knowledge is an input-independent, broad knowledge extracted from a pre-constructed knowledge graph. The specific knowledge is an input dependent, fine-grained knowledge obtained by retrieving reports with similar label distribution;
    \item We propose a novel knowledge-enhanced multi-head attention module to combine the structural information of knowledge and visual features; 
    \item Our experiments demonstrate consistent performance improvements when adding our proposed modules. Furthermore, our model achieves state-of-the-art performance on almost metrics for the public IU-Xray dataset and is on par with state-of-the-art methods for the MIMIC-CXR dataset.

\end{itemize} 

\begin{figure*}[!t]
\centerline{\includegraphics[width=0.95\linewidth]{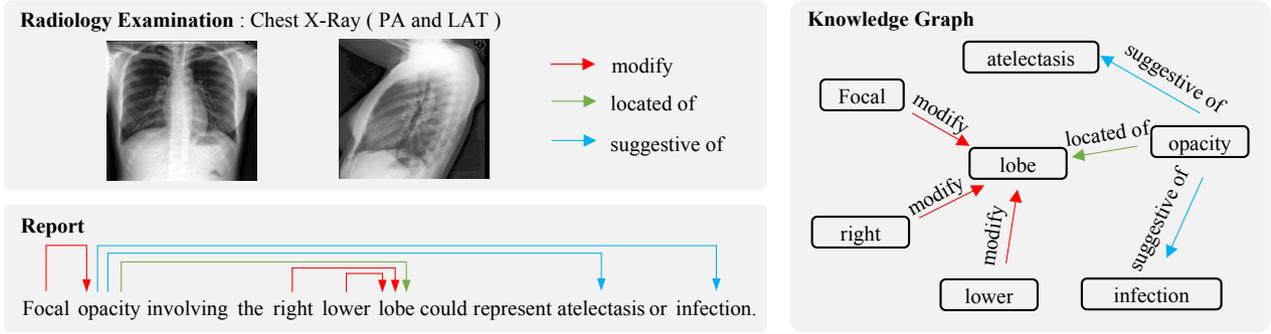}}
\caption{A radiology examination record and corresponding knowledge graph. The figure displays three parts of information. The first part includes posterior-anterior(PA) and lateral(LAT) chest x-ray images. The second part is the corresponding report which is highlighted by different relationships. The third part is the corresponding knowledge graph.}
\label{fig:graph}
\end{figure*}

\section{Related Works}
With the advancement of computer vision and natural language processing, many works have been exploited to combine radiology images and clinical information for automatically generating reports to assist radiologists~\citep{zhou2019handbook}. Inspired by image captioning, \cite{Shin2016Learning} adopt the CNN-RNN framework to describe the detected diseases based on visual features on a chest x-ray dataset. The work is restricted to the categories of pre-defined diseases. The Co-Attention(Co-Att)~\citep{Jing2018On} and \citep{Xue2018Multimodal,Yuan2019Automatic} propose different attention mechanisms and hierarchical LSTM to generate radiology reports. However, most reports generated by such works tend to describe normal observations, which indicates that such methods have difficulties capturing subtle changes in the image. TieNet~\citep{Wang2018TieNet} proposes an attention encoded text embedding and a saliency weighted global average pooling to boost the image classification and report generation. However, TieNet uses report embeddings as a part of LSTM's input which is not available in the inference stage, resulting in an embedding bias problem in the report generation. \cite{Liu2021Contrastive} propose a contrastive learning mechanism to capture and depict the changes by comparing the input radiology image with known normal radiology images. However, the performance of this method is restricted by the quality of the normal input images. \cite{Liu2021Competence} introduce a curriculum learning in medical report generation to help existing models better utilize the limited data and alleviate data bias. \cite{you2021aligntransformer} propose the AlignTransformer framework to alleviate the data bias problem and model the long sequence for radiology report generation. \cite{Nooralahzadeh2021Progressive} propose a two-stage radiology generation model by incorporating high-level concepts. Nevertheless, it needs extra labour to extract high-level context to guide the first stage generation.

Other types of work are also explored, which inject extra prior knowledge into the generation model to improve the quality of the generated radiology reports. Following the writing habit of radiologists, HRGR~\citep{Li2018Hybrid} compiles a manually extracted template database to generate radiology reports by reinforcement learning. The Knowledge-driven Encode, Retrieve Paraphrase(KERP)\citep{Li2019Knowledge}, the MKG~\citep{Zhang2020when} and the PPKED~\citep{Liu2021Exploring} propose to combine pre-constructed knowledge graph for radiology report generation. Although these methods achieved remarkable performance improvement, building the template database knowledge graph is still laborious, making it hard to transfer those approaches directly to other datasets. The Relational Report GENration(R2Gen)~\citep{Chen2020Generating} proposes a relational memory in the decoder to learn the order of words or sentences. However, the order is not essential as different radiologists have different writing habits. 

\begin{figure*}[!t]
\centerline{\includegraphics[width=0.95\linewidth]{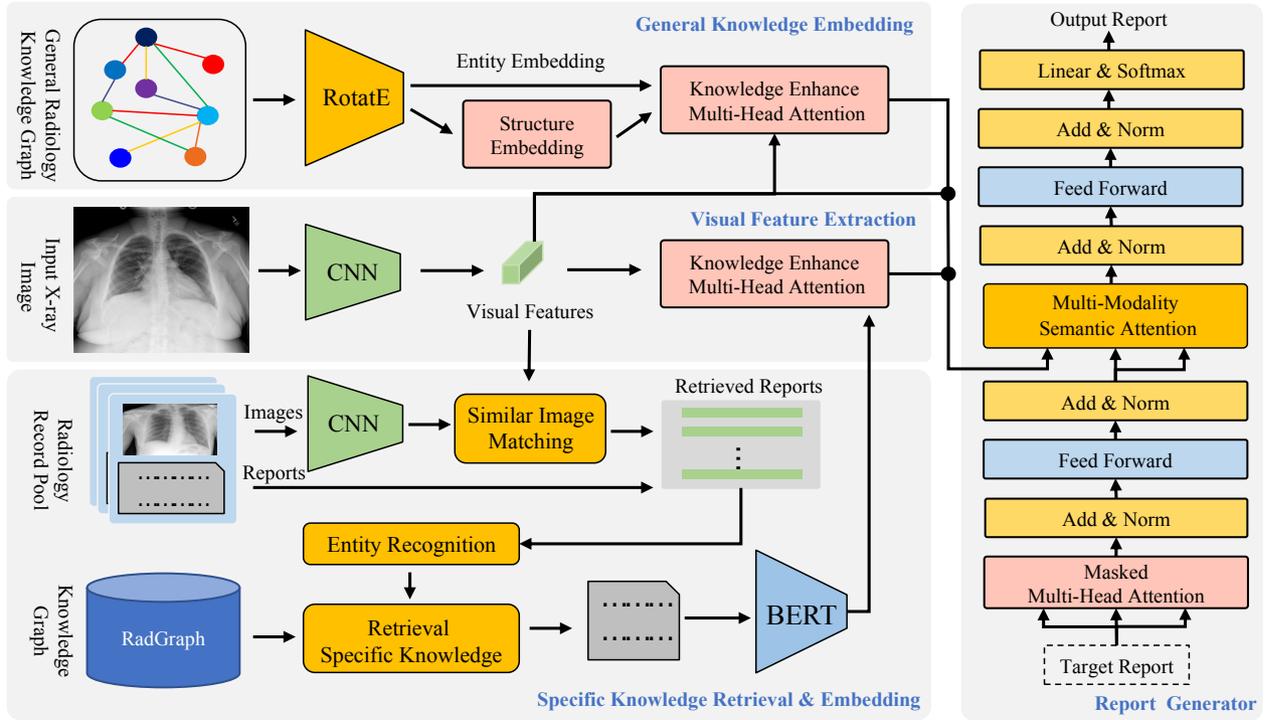}}
\caption{The proposed model architecture includes four major modules: the \textbf{general knowledge embedding} module embeds the general knowledge (GK) and attend the visual features by knowledge-enhanced multi-head attention; the \textbf{visual feature extraction} module extracts visual features and predicts the disease label's distribution from a single view or multiple views x-ray images by convolution neural network (CNN).  The \textbf{specific knowledge retrieval \& embedding} module retrieves the specific knowledge (SK) of the input x-ray image and integrates it in report generation. The \textbf{report generator} module fuses the general and specific knowledge to generate the radiology report. The \textbf{Knowledge Enhance Multi-Head Attention} is illustrated in Figure \ref{fig:KEMHA}.}
\label{fig:model}
\end{figure*}
\section{Method}
In this section, we introduce the proposed method that uses both general and specific knowledge for radiology report generation: In section~\ref{sec:overview}, we introduce the notation and provide an overview of the proposed method; In section~\ref{sec:external}, we describe how to integrate general knowledge in modelling and propose the knowledge-enhanced attention mechanism; In section~\ref{sec:internal}, we propose how to integrate the specific knowledge in modelling; In section~\ref{sec:generation}, we provide the multi-modality semantic attention module which combines both general and specific knowledge to generate the report.

\subsection{Overview}
\label{sec:overview}
Given a radiology x-ray image, we denote it as $Img \in \mathbb{R}^{C \times H \times W}$ where $C, H, W$ refers to the number of channels, height, and width of the input image, respectively. The radiology report generation task aims to generate a long sequence $W = \{w_1, \ldots, w_T\}$ to address the findings of this image, including both the normal and abnormal regions, where $w_t$ denotes the index of the word in the vocabulary $\mathbb{V}$ and $T$ denotes the number of words in the radiology report. 

The knowledge graph is a semantic network that reveals the relationship between entities. It can be regarded as a multi-relational graph, which contains many types of nodes and edges. The nodes usually represent different entities, and the edges usually represent different relationships in the knowledge graph. The entity refers to the nouns such as person, place, and the concept existing in the real world. The relationship usually represents some kind of connection between different entities. As shown in Fig~\ref{fig:graph}, a piece of knowledge in the knowledge graph of radiology report: \texttt{"opacity" - "suggestive of" - "atelectasis"}. In this knowledge, "opacity" and "atelectasis" are two specific concepts in the real world, and "suggestive of" indicates a relationship between the two concepts.

To include medical knowledge in modelling, we first define two types of medical knowledge: general knowledge and specific knowledge. For the general knowledge, we adopt the knowledge graph provided by RadGraph~\citep{Jain2021RadGraph}, which is manually built by board-certified radiologists from 500 radiology reports in the MIMIC-CXR data set. We use the set of triplets $\mathcal{G}_{g}= \{(source\ entity, relation, target\ entity)_i\}^{N_{g}}_{i=0}$ to denote general knowledge, where $N_{g}$ refers to the number of triplets. 

The acquired general knowledge includes the general medical knowledge about radiology reports. However, in the case of a specific radiology image, the general knowledge could be too broad to cover all details of the current input. Therefore, in addition to the general medical knowledge, we also propose to mine specific knowledge for each input image. Given the input image $Img^{(t)}$, we retrieve similar reports from a pre-built report repository according to the visual appearance. Then we apply the relation extractor provided by~\citep{Jain2021RadGraph} to acquire triplets from these retrieved reports. We regard the acquired triplets $\mathcal{G}_{s}^{(t)} = \{(source\ entity, relation, target\ entity)_i^{(t)}\}^{N_{s}}_{i=0}$ as the specific knowledge for $Img^{(t)}$, where $N_{s}$ denote the number of triplets.

Figure~\ref{fig:model} displays the overflow of the proposed model, which includes four major modules:

The \textbf{visual feature extraction (VFE)} module extracts visual features $I$. It predicts the disease label's distribution $Y$ from a single x-ray image or multiple images of different views by a convolution neural network (CNN), e.g. $I = CNN(Img),  I \in \mathbb{R}^{k \times d}$, where $k$ and $d$ refer to the number of visual feature channels and the dimension of visual features. 
    
The \textbf{General Knowledge embedding (GKE)} module embeds the General Knowledge $\mathcal{G}_{g}$ to acquire entity embeddings and relation embeddings via RotatE model~\citep{Sun2019Rotate} and attend the visual features by knowledge-enhanced multi-head attention. The details will be discussed in section~\ref{sec:external}. 

The \textbf{specific knowledge retrieval \& embedding (SKE)} module retrieves the specific knowledge $\mathcal{G}_{s}^{(t)}$ of current input x-ray image and integrates it in report generation. The details will be disused in section~\ref{sec:internal}.

The \textbf{report generator} module fuses the general and specific knowledge to generate the radiology report. The details will be discussed in section~\ref{sec:generation}.

\subsection{General Knowledge Embedding}
\label{sec:external}
In this subsection, we describe how to integrate general medical knowledge in generating radiology reports. The general medical knowledge such as \textit{effusion located at pleural} and \textit{atelectasis suggestive of consolidation}, if used properly, can help generate a more accurate report. Here, we use the knowledge graph provided by RadGraph~\citep{Jain2021RadGraph} as the general knowledge. This knowledge graph is manually built by board-certified radiologists and represents the radiology knowledge in triplets \textit{(source entity, relation, target entity)}.  

\subsubsection{Embedding the General Knowledge}
To derive knowledge from this manual built knowledge graph, we employ a graph embedding model RotatE~\citep{Sun2019Rotate} to acquire entity embeddings and relation embeddings by Eq.(\ref{eq:rotate}). The RotatE model defines each relation as a rotation from the source entity to the target entity in the complex vector space.
\begin{align}
    \{E_e, E_r\} = RotatE(\mathcal{G}_{g}), \label{eq:rotate}
\end{align}
where $E_e \in \mathbb{R}^{N_e \times 400}$ and $E_r \in \mathbb{R}^{N_r \times 400}$ denote the entity embeddings and relation embeddings, respectively. $N_e$ and $N_r$ refer to the number of entity and relation in  general knowledge, respectively.

\subsubsection{Knowledge-Enhanced Multi-Head Attention}

\begin{figure}[!t]
\centerline{\includegraphics[width=\columnwidth]{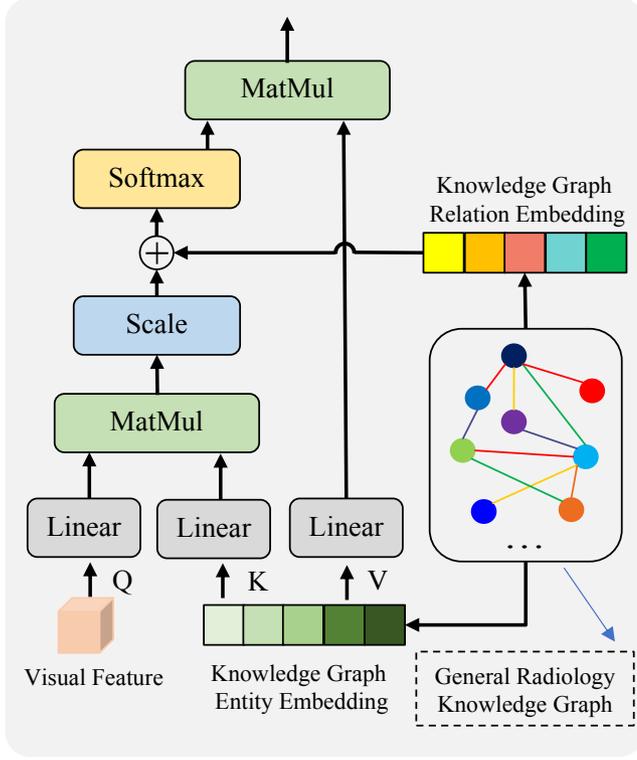}}
\caption{An illustration of Knowledge-Enhanced Multi-Head Attention. It combines entity and relation embeddings of general radiology knowledge graph in multi-head attention.}
\label{fig:KEMHA}
\end{figure}

Once we acquire the entity embedding $E_e$ and relation embedding $E_r$, the next step is to combine $E_e$ and $E_r$ with the visual feature $I$. Previous approaches \citep{Li2018Hybrid, Zhang2020when} only use entity embedding and calculate the semantic correlation between visual feature $I$ and entity embedding $E_e$. As a result, the structural information $E_e$ of the knowledge graph is missed in modelling. To include the structural information, \citep{hu2020open,gilmer2017neural,xu2018powerful} aggregates the edge features over the associated node and combines the aggregated edge features with the node feature in modelling; The Graphormer~\citep{Ying2021Do} proposes to encode the edge features of the shortest path as inductive bias. Inspired by previous works, for each entity in the general knowledge, we propose aggregating the edge features of all its neighbours and adding it as a relation bias in modelling.

Firstly, we build aggregate relation embedding $\mathbf{r} \in \mathbb{R}^{N_e \times N_e \times 400}$ from the general knowledge where each element $\mathbf{r}_{ij} \in \mathbb{R}^{400 \times 1}$ represents the relations between the source entity $e_i$ and the target entity $e_j$. Due to more than one relations between the two entities, we use the average of the relation embeddings as $\mathbf{r}_{ij}$ by Eq.(\ref{eqn:relation}).
\begin{align}
    \mathbf{r}_{ij} &= AvgPool(E_r^{(i,j)}), \label{eqn:relation} \\
    E_r^{(i,j)} &= Ralations(e_i, e_j), \nonumber
\end{align} 
where $E_r$ is acquired in Eq.(\ref{eq:rotate}), $AvgPool(\cdot)$ refers to average pooling function, and $Ralations(e_i, e_j)$ refers to all relations between the entity $e_i$ and the entity $e_j$.

Secondly, the relation bias is acquired by Eq.(\ref{eqn:structure}):
\begin{align}
    \phi(\mathbf{r}) = AvgPool(\mathbf{r}W^R), \label{eqn:structure}
\end{align}
where $W^R$ are learnable parameters of linear projection in the size of $400\times 1$ and $\mathbf{r}W^R$ is in the size of $N_e \times N_e$. After pooling, $\phi(\mathbf{r}) \in \mathbb{R}^{1 \times N_e}$ is the aggregated relation embedding of entities.

Following the Graphormer~\citep{Ying2021Do} and the multi-head attention~\citep{Vaswani2017Attention}, we propose a knowledge-enhanced attention mechanism which is formulated as follows:
\begin{align}
    KG\text{-}Att(\mathbf{Q, K, V}, \phi(\mathbf{r})) = softmax(\frac{\mathbf{QK}^T}{\sqrt{d_k}} + \phi(\mathbf{r}))\mathbf{V},
\end{align}
where $\mathbf{Q}$, $\mathbf{K}$ and $\mathbf{V}$ denote the query, key and value features. $d_k$ denotes the dimension of embedding.

By applying the multi-head attention to the \textit{KG\text{-}Att} function, the knowledge-enhanced multi-head attention(KEMHA) is defined in Eq.(\ref{eq:kgmha}). The Figure \ref{fig:KEMHA} shows the architecture of KEMHA. 
\begin{align}
    KEMHA(\mathbf{Q}, &\mathbf{K, V}, \mathbf{r}) = [head_1, \ldots, head_h]W^O, \label{eq:kgmha} \\
    head_i &= KG\text{-}Att(\mathbf{Q}W^Q_i, \mathbf{K}W^K_i, \mathbf{V}W^V_i, \phi_i(\mathbf{r})),
    \nonumber
\end{align}
where $\phi_i(\mathbf{r})=AvgPool(\mathbf{r}W_i^R)$ denotes the relation bias for the $i^{th}$ head, $W^Q_i, W^K_i, W^V_i, W^R_i$, and $W^O$ are learnable parameters of linear projection. 

Finally, the general knowledge augmented visual representations are acquired by Eq.(\ref{eq:ve}). Here we include both the entity embedding $E_e$ and the structural information $\mathbf{r}$.
\begin{align}
    C_g &= KEMHA(I, E_e, E_e, \mathbf{r}). \label{eq:ve}
\end{align}
Also, we apply the layer normalization on all embeddings to normalize them to the same scale.

\subsection{Specific Knowledge Retrieval \& Embedding}
\label{sec:internal}
The general knowledge introduced in the previous subsection provides the basic medical knowledge for radiology report generation. However, the general knowledge could be too broad to cover all the details for a specific input image. Therefore, we also retrieve the specific knowledge that is closely related to the current input for each input image. Here we propose a two-step approach: 1) Retrieve similar reports from a pre-built repository; 2) Mine specific knowledge from retrieved reports.  
\subsubsection{Similar Record Retrieval}
Here we assume that given an input image, similar historical records could be used as references in report generation. Therefore, we firstly build a radiology record repository $\{Img^{(i)}, R^{(i)}\}_{i=0}^{N_r}$. In the context of this paper, we directly use the training data set as the repository. In this manner, there is no overlap between retrieved reports and reports in the testing set. The label leakage is avoided. Given an input radiology image, we search for similar records from this pre-built repository. The details are as follows:

For each record in the repository, we firstly extract visual feature $I^{(i)} = CNN(Img^{(i)})$ and then calculate the distribution over disease labels with $I^{(i)}$. 

\begin{align}
   Y^{(i)} = softmax(AvgPool(I^{(i)})W_c + b_c), \label{eq:label} 
\end{align}
where $Y^{(i)} \in \mathbb{R}^{1 \times N_c}$, $N_c=14$ refers to the number of disease labels, and $W_c$ and $b_c$ are the weight parameter and bias term of a linear transform for disease classification, respectively. 

We calculate the KL-Divergence between the disease distribution $Y$ of the input image and the disease distribution $\{Y^{(i)}\}^{N_{r}}_{i=0}$ of samples in the repository in Eq. (\ref{eqn:score}). The acquired score is used to select the most relevant candidates from the repository:
\begin{align}
    score = 1 - \sum_{j=0}^{N_c}Y_{j}ln\frac{Y_{j}}{Y_{j}^{(i)}}.
    \label{eqn:score}
\end{align}

\subsubsection{Mining Specific Knowledge from Similar Records}
For each input image $Img$, after acquiring the top-k most similar records $\{Img^{(i)}, R^{(i)}\}_{i=0}^{k}$ from the repository, we extract the named entities $T^{(i)}$ from each retrieved report $R^{(i)}$. Here we use Stanza, a named entity recognizer provided by \citep{qi2020stanza,Zhang2021Biomedical}. For example, given a retrieved report $R^{(i)}$ of  ``\textit{pneumothorax or pleural effusion is seen}", the set of extracted entities $T^{(i)}$ is \{\textit{pneumothorax, pleural, effusion}\}.
\begin{align}
    T^{(i)} = \{t_1^{(i)}, t_2^{(i)},\ldots \}.
\end{align}

Once we acquire entities from similar reports, we use these entities to query specific knowledge. In detail, we first obtain the set of triplets from RadGraph~\citep{Jain2021RadGraph}. These triplets are automatically extracted from the all training data of MIMIC-CXR dataset by label tools\citep{Jain2021RadGraph} which includes 6M entities and 4M relations. The volume of this triplet set is 40 times larger than the one used in Section \ref{sec:external} which only use the part of manually built by board-certified radiologists from 500 radiology reports in the MIMIC-CXR dataset, thus including much more knowledge.
For example, when the entities $T^{(i)}$ is \{\textit{pneumothorax, effusion}\}, the retrieved triplets
$\mathcal{K} ^{(i)}$ are \{\textit{pneumothorax suggestive of bleeding, effusion located at bilateral},...\}.
\begin{align}
    \mathcal{K} ^{(i)} &= \{k^{(i)}_1, k^{(i)}_2, \ldots\},
\end{align}
where $k_j^{(i)}$ denotes the triplet retrieved via matching radiology term entity $t_j^{(i)}$.

Then for each image, we concatenate all acquired triplets into one single sentence and feed this sentence to ClinicalBERT~\citep{alsentzer2019publicly}. The pooled outputs of clinical BERT is denoted as $E_s$.
\begin{align}
    E_{s} = BERT(\mathcal{K} ),
\end{align}
here $E_s$ can be viewed as the specific knowledge for each input image. We further fuse the specific knowledge into visual features by multi-modality semantic attention.
The multi-modality context representations $C_{s}$ are acquired by Eq.(\ref{eq:mmsa}):
\begin{align}
    C_{s} = KEMHA(I, E_{s}, E_{s}, \mathbf{0}), \label{eq:mmsa}
\end{align}
where \textit{KEMHA} denotes the knowledge-enhanced multi-Head attention that is defined in Eq.(\ref{eq:kgmha}). Since the specific knowledge is organized as a sequence of facts and the structural information is embedded into the sequence,  we set the structural bias to zero $\mathbf{0}$.

\subsection{Radiology Report Generator}
\label{sec:generation}

We employ the standard decoder of Transformer~\citep{Vaswani2017Attention} to generate radiology reports via concatenating the extracted visual features, general knowledge, and specific knowledge.
\begin{align}
    w_t = Decoder(concat&(I, C_{g}, C_{s}), w_{<t}), w_0 = \mathbf{0}. \label{eq:decoder}
\end{align}

The generator follows the auto-regressive decoding process generating the medical report sequence $\mathbf{W}$ from the conditional distribution:
\begin{align}
    p(\mathbf{W}|Img,\mathcal{G}_{s}, \mathcal{G}_{g}) = &\prod_{t=0}^{T} p_\theta(w_t|w_{<t},Img,\mathcal{G}_{s}, \mathcal{G}_{g}).
\end{align}

Finally, the whole model is optimized by the negative conditional log-likelihood of $p(\mathbf{W})$ given the image and its general/specific knowledge:
\begin{align}
    \mathcal{L} = \sum^{T}_{t=0} log\ p_\theta(w_t|w_{<t},Img,\mathcal{G}_{s}, \mathcal{G}_{g}).
\end{align}

\section{Experiments and Analysis}
\subsection{Datasets and Baselines}
\label{sec:dataset}
\begin{table*}[htbp]
  \centering
  \caption{The performances of our model compared with baselines on IU-Xray dataset. The best results are highlighted in bold. For the baselines marked by *, we cite the results reported in \citep{Jing2019Show}. For the baselines marked by $\#$, we replicate results by their codes; the rests are cited from the original paper. BLEU-n measures the accuracy, and ROUGE-L measures the recall of the generated report. CIDEr evaluates whether the generated report by key information in the ground truth report. The higher the values means the better the performance of the model.}
    \begin{tabular}{ccccccc}
    \toprule
    Model & BLEU-1 & BLEU-2 & BLEU-3 & BLEU-4 & CIDE-r & ROUGE-L \\
    \midrule
    S\&T*~\citep{vinyals2015show}  & 0.216  & 0.124  & 0.087  & 0.066  & 0.294  & 0.306  \\
    SA\&T$^\#$~\citep{xu2015show} & 0.399  & 0.251  & 0.168  & 0.118  & 0.302  & 0.323  \\
    AdaAtt*~\citep{lu2017knowing} & 0.220  & 0.127  & 0.089  & 0.068  & 0.295  & 0.308  \\
    CMAS*~\citep{Jing2019Show}  & 0.464  & 0.301  & 0.210  & 0.154  & 0.275  & 0.362  \\
    CoAtt*~\citep{Jing2018On} & 0.455  & 0.288  & 0.205  & 0.154  & 0.277  & 0.369  \\
    HRGR~\citep{Li2018Hybrid}  & 0.438  & 0.298  & 0.208  & 0.151  & 0.343  & 0.322  \\
    KERP~\citep{Li2019Knowledge}  & 0.482  & 0.325  & 0.226  & 0.162  & 0.280  & 0.339  \\
    R2Gen~\citep{Chen2020Generating} & 0.470  & 0.304  & 0.219  & 0.165  & /     & 0.371  \\
    PPKED~\citep{Liu2021Exploring} & 0.483  & 0.315  & 0.224  & 0.168  & 0.351  & 0.376  \\
    CA~\citep{Liu2021Contrastive}    & 0.492  & 0.314  & 0.222  & 0.169  & /     & 0.381  \\
    CMCL~\citep{Liu2021Competence}  & 0.473  & 0.305  & 0.217  & 0.162  & /     & 0.378  \\
    AlignTransformer~\citep{you2021aligntransformer} & 0.484  & 0.313  & 0.225  & 0.173  & /     & 0.379  \\
    M2TR~\citep{Nooralahzadeh2021Progressive}  & 0.486  & 0.317  & 0.232  & 0.173  & /     & \textbf{0.390 } \\
    \midrule
    Ours  & \textbf{0.496 } & \textbf{0.327 } & \textbf{0.238 } & \textbf{0.178 } & \textbf{0.382 } & 0.381  \\
    \bottomrule
    \end{tabular}%
  \label{tab:NLG-iu}%
\end{table*}%

\textbf{MIMIC-CXR.} MIMIC-CXR~\citep{Johnson2019MIMIC-CXR-JPG} is a large-scale labeled chest radiology dataset that includes 377,110 x-ray images and corresponding 227,827 radiology studies for 65,379 patients. The standard dataset split is adopted. We preprocess each report by converting all tokens to lowercase and filtering the tokens with no more than three occurrences, resulting in 5290 unique words.

\textbf{IU-Xray.} Indiana University Chest X-ray Collection (IU-Xray)~\citep{Demner-Fushman2016iu-dataset} is a public radiology examination dataset and broadly used to evaluate the performance in previous works of this task. The dataset includes 7,470 x-ray images and corresponding 3,955 radiology reports. There is no standard dataset split of the IU-Xray dataset. To fairly compare with previous work, the dataset split provided by R2Gen~\citep{Chen2020Generating} is adopted that randomly splits the data into training, validation, and testing sets with a ratio of 7:1:2 without overlap in patients. We use the same preprocessing mentioned above for the IU-Xray dataset, which results in 855 unique words. 

The labels of both datasets include 12 diseases and 2 indication labels such as enlarged cardiomediastinum, cardiomegaly, lung lesion, airspace opacity, edema, consolidation, pneumonia, atelectasis, pneumothorax, pleural effusion, pleural other, fracture, support devices, and no finding.

A knowledge graph is a large-scale graph connected by semantic relations among entities. Each semantic relation is composed of a knowledge triplet \textit{\{source entity, relation, target entity\}}. The RadGraph~\citep{Jain2021RadGraph} is a knowledge graph of clinic radiology entities and relations based on full-text chest x-ray radiology reports.It is worth mentioning that we build our general and specific knowledge on the RadGraph only on the training set of the MIMIC-CXR dataset. Thus there is no information leakage.

\textbf{General Knowledge.}
The general knowledge is a knowledge graph that incorporates extra radiology knowledge into our model. The general knowledge is extracted from RadGraph's development dataset, which is annotated by board-certified radiologists. RadGraph contains 500 radiology reports sampled from the MIMIC-CXR dataset, resulting in 14,579 entities and 10,889 relations. We filter the entities that occur more than ten times and corresponding relations as general knowledge resulting in 239 entities, three relations, and 1,508 knowledge triplets.

\textbf{Specific Knowledge.}
The specific knowledge is extracted from RadGraph's inference dataset, which is automatically annotated by 220,763 reports from the MIMIC-CXR dataset by DYGIE++~\citep{wadden2019entity} and Bio+Clinical BERT~\citep{alsentzer2019publicly} model. The inference dataset results in 6,161,943 entities and 4,409,026 knowledge triplets. We filter the entities that occur more than three times and remove the redundancy triplets. The entities and knowledge triplets extracted from training reports of MIMIC-CXR are selected as our specific knowledge, including 7,093 entities and 67,722 triplets.

\textbf{Baselines.}
We compare our proposed model with general image captioning works, e.g. \textbf{S\&T}~\citep{vinyals2015show}, \textbf{SA\&T}~\citep{xu2015show}, \textbf{AdaAtt}~\citep{lu2017knowing},  \textbf{TopDown}~\citep{anderson2018bottom}, and chest radiology report generation works, e.g. \textbf{CoAtt}~\citep{Jing2018On}, \textbf{CMAS}~\citep{Jing2019Show}, \textbf{HRGR}~\citep{Li2018Hybrid}, \textbf{KERP}~\citep{Li2019Knowledge} \textbf{PPKED}~\citep{Liu2021Exploring}, \textbf{R2Gen}~\citep{Chen2020Generating}, \textbf{CA}~\citep{Liu2021Contrastive}, \textbf{CMCL}~\citep{Liu2021Competence}, \textbf{AlignTransformer}~\citep{you2021aligntransformer}, and \textbf{M2TR}~\citep{Nooralahzadeh2021Progressive}.

\begin{table*}[htbp]
  \centering
  \caption{The performances of our model compared with baselines on MIMIC-CXR dataset. The best results are highlighted in bold. For the baselines marked by $\#$, we replicate results by their codes; the rests are cited from the original paper. BLEU-n measures the accuracy, and ROUGE-L measures the recall of the generated report. CIDEr evaluates whether the generated report by key information in the ground truth report. The higher the values means the better the performance of the model.}
    \begin{tabular}{ccccccc}
    \toprule
    Model & BLEU-1 & BLEU-2 & BLEU-3 & BLEU-4 & CIDE-r & ROUGE-L \\
    \midrule
    S\&T$^\#$~\citep{vinyals2015show}  & 0.256  & 0.157  & 0.102  & 0.070  & 0.063  & 0.249  \\
    SA\&T$^\#$~\citep{xu2015show} & 0.304  & 0.177  & 0.112  & 0.077  & 0.083  & 0.249  \\
    AdaAtt$^\#$~\citep{lu2017knowing} & 0.311  & 0.178  & 0.111  & 0.075  & 0.084  & 0.246  \\
    TopDown$^\#$~\citep{anderson2018bottom} & 0.280  & 0.169  & 0.108  & 0.074  & 0.073  & 0.250  \\
    R2Gen~\citep{Chen2020Generating} & 0.353  & 0.218  & 0.145  & 0.103  & /     & 0.277  \\
    PPKED~\citep{Liu2021Exploring} & 0.360  & 0.224  & 0.149  & 0.106  & /     & \textbf{0.284 } \\
    CA~\citep{Liu2021Contrastive}    & 0.350  & 0.219  & 0.152  & 0.109  & /     & 0.283  \\
    CMCL~\citep{Liu2021Competence}  & 0.334  & 0.217  & 0.140  & 0.097  & /     & 0.281  \\
    AlignTransformer~\citep{you2021aligntransformer}    & \textbf{0.378 } & \textbf{0.235 } & 0.156  & 0.112  & /     & 0.283  \\
    M2TR~\citep{Nooralahzadeh2021Progressive}  & \textbf{0.378 } & 0.232  & 0.154  & 0.107  & /     & 0.272  \\
    \midrule
    Ours  & 0.363  & 0.228  & \textbf{0.156 } & \textbf{0.115 } & \textbf{0.203 } & \textbf{0.284}  \\
    \bottomrule
    \end{tabular}%
  \label{tab:NLG-mimic}%
\end{table*}%

\begin{table}[htbp]
  \centering
  \caption{The results of clinical efﬁcacy (CE) metrics on the MIMIC-CXR dataset. The best results are highlighted in bold. For the baselines marked by $^\#$, we replicate results by their codes; the rests are cited from the original paper.}
    \resizebox{0.48\textwidth}{!}{
    \begin{tabular}{c|cccc}
    \toprule
    Model & Accuracy & Precision & Recall & F1-Score \\
    \midrule
    S\&T$^\#$  & 0.423  & 0.084  & 0.066  & 0.072  \\
    SA\&T$^\#$ & 0.703  & 0.181  & 0.134  & 0.144  \\
    AdaAtt$^\#$ & 0.741  & 0.265  & 0.178  & 0.197  \\
    TopDown$^\#$ & 0.743  & 0.166  & 0.121  & 0.133  \\
    M2TR  & /     & 0.240  & \textbf{0.428}  & 0.308  \\
    VTI   & /     & 0.350  & 0.151  & 0.210  \\
    R2Gen & /     & 0.333  & 0.273  & 0.276  \\
    \midrule
    Ours wo/KEMHA & 0.792  & 0.413  & 0.286  & 0.317  \\
    Ours wo/GKE & 0.815  & 0.451  & 0.345  & 0.368  \\
    Ours wo/SKE & 0.810  & 0.433  & 0.331  & 0.352  \\
    Ours  & \textbf{0.816 } & \textbf{0.458 } & 0.348  & \textbf{0.371 } \\
    \bottomrule
    \end{tabular}%
    }
  \label{tab:CE}%
\end{table}%

\begin{table*}[htbp]
  \centering
  \caption{The performances of our model compared with the model without different proposed modules on IU-Xray and MIMIC-CXR datasets. The Base model refers to a model that removes all proposed modules. The best results are highlighted in bold. The \textit{wo} is the abbreviation of \textit{without}.}
    \resizebox{0.78\textwidth}{!}{
    \begin{tabular}{c|c|cccccc}
    \toprule
    Dataset & \multicolumn{1}{c|}{Model} & BLEU-1 & BLEU-2 & BLEU-3 & BLEU-4 & CIDE-r & ROUGE-L \\
    \midrule
    \multirow{5}[4]{*}{IU-Xray} & Base  & 0.357  & 0.221  & 0.158  & 0.121  & 0.325  & 0.149  \\
          & Ours wo/KEMHA & 0.401  & 0.258  & 0.186  & 0.143  & 0.348  & 0.167  \\
          & Ours wo/GKE & 0.422  & 0.274  & 0.202  & 0.159  & 0.358  & 0.176  \\
          & Ours wo/SKE & 0.447  & 0.286  & 0.205  & 0.157  & 0.364  & 0.187  \\
\cmidrule{2-8}          & Ours  & \textbf{0.496 } & \textbf{0.327 } & \textbf{0.238 } & \textbf{0.178 } & \textbf{0.381 } & \textbf{0.200 } \\
    \midrule
    \multirow{5}[4]{*}{MIMIC\_CXR} & Base  & 0.333  & 0.207  & 0.141  & 0.103  & 0.276  & 0.132  \\
          & Ours wo/KEMHA & 0.358  & 0.222  & 0.151  & 0.110  & 0.279  & 0.139  \\
          & Ours wo/GKE & 0.358  & 0.224  & 0.153  & 0.112  & 0.283  & 0.142  \\
          & Ours wo/SKE & 0.354  & 0.222  & 0.152  & 0.110  & 0.282  & 0.141  \\
\cmidrule{2-8}          & Ours  & \textbf{0.363 } & \textbf{0.228 } & \textbf{0.156 } & \textbf{0.115 } & \textbf{0.284 } & \textbf{0.144 } \\
    \bottomrule
    \end{tabular}%
    }
  \label{tab:ablation}%
\end{table*}%

\subsection{Experimental Setup and Metrics}
In our implementation, all images are resized to 224$\times$224. The ResNet-101~\citep{He2016deep} model pre-trained on ImageNet is employed as our visual encoder, and the pre-trained ClinicalBERT~\citep{alsentzer2019publicly} model is used as a specific knowledge encoder. The ResNet-101 model pre-trained on MIMIC-CXR dataset by torchXrayVision~\citep{Cohen2021TorchXRayVision} is employed as the feature extractor for retrieving similar images. The parameters of the decoder are randomly initialized, and the number of layers is set to 3. The dimension of the model is set to 512. We use a fixed mini-batch and zero-padding for each mini-batch report to keep the same length. We train our model 50/30 epochs for the IU-Xray/MIMIC-CXR dataset, respectively. We adopt the Adam optimizer with an initial learning rate of 1e-4 and weight decay of 5e-5 for training except for visual encoder initial with a learning rate set to 5e-5. We acquire top-10 similar records for extracting specific knowledge. All hyper-parameters are determined by the performance of the BLEU-4 score on the validation dataset, and the corresponding testing dataset results are reported.

\textbf{Metrics.}
Following other radiology report generation works, we evaluate the performance of the proposed method by the widely used natural language generation (NLG) metrics, such as BLEU-n~\citep{papineni2002bleu}, CIDEr~\citep{vedantam2015cider}, METEOR~\citep{denkowski2014meteor}, and ROUGE-L~\citep{lin2004rouge} score. The BLEU-n is used to measure accuracy of the generated report. It is a widely adopted machine translation metric that analyzes the co-occurrences of n-grams between the generated sentences and ground truth. The higher degree of co-occurrences means the higher quality of the generated text. The ROUGE-L is measured similar to BLEU-n, and the difference is that BLEU-n calculates the accuracy and ROUGE-L calculates the recall. The CIDEr evaluates whether the text generated by the model covers the key information in the ground truth. It regards each sentence as a document, then calculates the cosine of the TF-IDF vector, and compares the similarity between the generated text and the ground truth text in the vector space. The results are computed by MS-COCO caption evaluation tool~\footnote{https://github.com/tylin/coco-caption} automatically.

It is worth mentioning that the scores of traditional NLP metrics such as BLEU-n and ROUGE only depend on a tiny portion of words. They may not provide comprehensive assessments in terms of medical diagnosis. Therefore, we adopt clinical efficacy (CE) metrics~\citep{Chen2020Generating} to analyze the quality of generated reports from the perspective of clinics. ~\citep{Irvin2019CheXpert} proposes a tool named CheXpert labeler to extract medical terminology. The providers of MIMIC-CXR dataset use it to build the labels. Similar to the building process, the clinical efficacy metric uses the CheXpert labeler to extract labels from the ground truth and the generated reports automatically. The clinical efficacy (CE) metrics are defined by calculating the precision/recall/f1 scores between extracted labels from generated reports and ground-truth reports. This CE metric can help to evaluate how well the generated report describes the abnormalities. Because the IU-Xray dataset does not provide consistent labels, we only report CE metrics on the MIMIC-CXR dataset.

\subsection{Quantitative Results}

Table~\ref{tab:NLG-iu} and ~\ref{tab:NLG-mimic} show the experimental results of the baselines model and our proposed model evaluated by natural language generation (NLG) metrics on both datasets, respectively. There are four observations. Firstly, our proposed model outperforms almost all other radiology report generation models. Compared to the state-of-the-art model, our model has a performance improvement , leading to an increase of BLEU-4 from 0.173 to 0.178 on the IU-Xray dataset and from 0.107 to 0.115 on the MIMIC-CXR dataset. It demonstrates the effectiveness of incorporating general and specific knowledge and the knowledge enhanced method. Secondly, compared with general image captioning methods (Table~\ref{tab:NLG-iu} and ~\ref{tab:NLG-mimic}, row 1/2/3), the radiology report generation models (the rest of Table~\ref{tab:NLG-iu} and ~\ref{tab:NLG-mimic}) achieve remarkable improvement. It indicates that it is crucial to propose domain-specific techniques for radiology report generation. Thirdly, our proposed method outperforms all the previous models on the IU-Xray dataset over all metrics, except that  the ROUGE-L metric is slightly lower compared to M2TR~\citep{Nooralahzadeh2021Progressive}. The reason may be due to that the ROUGE-L focuses more on the recall of generated reports, and M2TR~\citep{Nooralahzadeh2021Progressive} proposes two-staged generation progress that first identifies accurate labels and later enriches those initial sequences to more fluent sentences. As a result, M2TR generates more description patterns in radiology reports, resulting a slightly higher ROUGE-L score. On the MIMIC-CXR dataset, our proposed method has higher performance than most of the previous methods. Although it is slightly lower than M2TR~\citep{Nooralahzadeh2021Progressive} and AlignTransformer~\citep{you2021aligntransformer} on the BLEU-1/2 scores, it achieves higher performances in terms of BLEU-3/4 scores.

Table~\ref{tab:CE} presents the experimental results of the baseline model and our proposed method evaluated by clinical efficacy (CE) metrics on the MIMIC-CXR dataset. The metrics are calculated by comparing the critical radiology terminology extracted from the generated and reference reports. Our method significantly outperforms the previous models. Compared with the state-of-the-art model R2Gen, our model has a performance boost, leading to an increase of 37.5\% on Precision, 27.5\% on Recall, and 34.4\% on F1-Score. The performance improvement demonstrates that our model can predict more accurate and critical information. 

\begin{figure*}[!t]
\centerline{\includegraphics[width=0.95\linewidth]{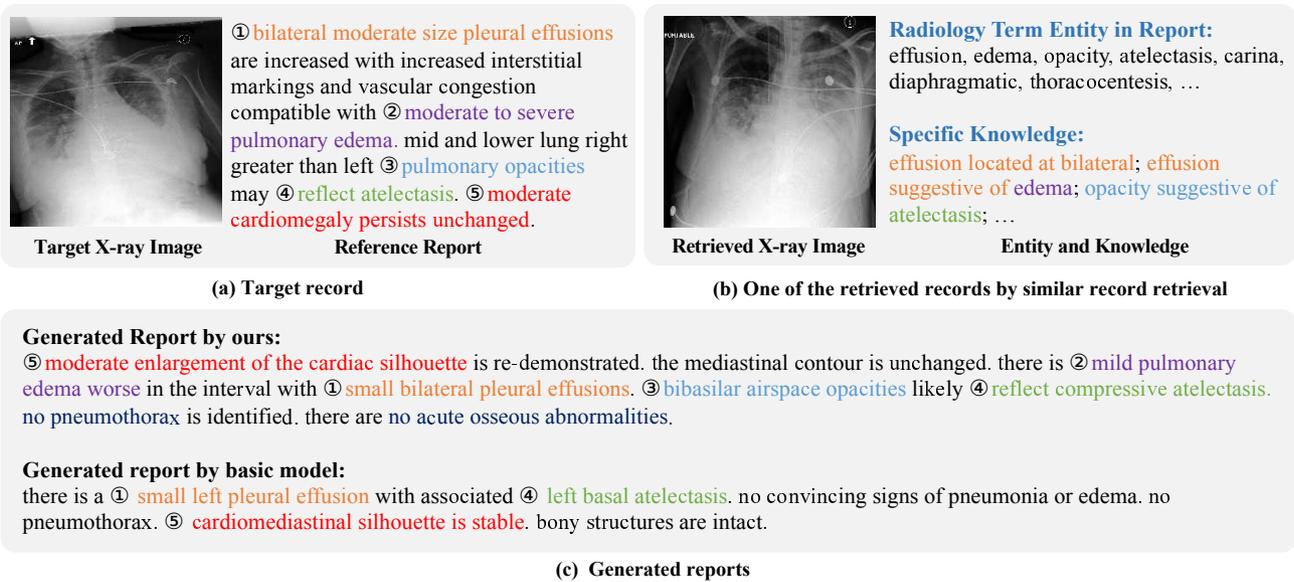}}
\caption{The reports were generated by our proposed method and basic model on the MIMIC-CXR dataset. The sub-figure (a) is the target record, including an x-ray image and a reference report. The sub-figure (b) is one of the retrieved records by our proposed similar record retrieval method that includes an x-ray image, a radiology terminology set which is extracted from the corresponding report by Stanza~\citep{qi2020stanza}, and the specific knowledge corresponding to the radiology terminologies. The sub-figure (c) is the visualization of generated reports by our proposed method and basic model. Our generated report preserves all the five types radiology terms(\textcircled{1} to \textcircled{5}) while the basic model only contains three of the five (\textcircled{1}, \textcircled{4} and \textcircled{5}).  In all sub-figures, the same radiology terminology is highlighted by the same colour.}
\label{Fig:case}
\end{figure*}

\subsection{Ablation Study}

To evaluate the effectiveness of each module, we conduct ablation studies on IU-Xray and MIMIC-CXR datasets. We adopt the vanilla transformer model serves as our base model. We remove each proposed module and maintain other parts in the subsequent studies. The results are shown in Table~\ref{tab:CE} and~\ref{tab:ablation}. 

\textbf{Removing knowledge-enhanced multi-head attention (KEMHA) module}. Unlike image or text mode, which only has one modality data, the knowledge graph consists of entity and relation. In previous knowledge-based works~\citep{Zhang2020when, Liu2021Exploring}, they only incorporate the entity embeddings into their models. However, the relation also contains important information. The knowledge-enhanced multi-head attention aggregates relation embeddings between entities. In this study, we replace the KEMHA module with a regular multi-head attention module which, same as previous works, only incorporates entity embeddings. As shown in the two Tables, when removing the KEMHA module, our model has a significant performance drop on both IU-Xray and MIMIC-CXR datasets. The results demonstrate that the relationship is crucial and suggest that radiology report generation can benefit from aggregating relation embeddings.

\textbf{Removing general knowledge embedding (GKE) module}. The general knowledge can be regarded as standard medical terms in radiology examinations. The model that without using the general knowledge removes the $C_{g}$ in Eq.(\ref{eq:decoder}). The two tables consistently show that the model without general knowledge is relatively lower than the original model. It demonstrates the effectiveness of the general knowledge and suggests that standard medical terms can guide the model in learning more fine-grained features from visual features. Besides, we observe that the model without GKE outperforms the model without KEMHA. But, compared with the model without GKE, the model without KEMHA incorporates extra knowledge, for example entity embeddings, which indicates that the incomplete knowledge graph may add a bias into the model and lead to performance degeneration.

\textbf{Removing specific knowledge embedding (SKE) module}. The specific knowledge can be regarded as retrieving a few target-related knowledge from the knowledge pool. The model that without using the specific knowledge does not need to retrieve related radiology records and removes the $C_{s}$ in Eq.(\ref{eq:decoder}).  The model without SK is relatively lower than the original model, demonstrating the effectiveness of the specific knowledge. 

\textbf{Impact of the knowledge graph embedding model.} We conducted an ablation study to analyze the impact of different the knowledge graph embedding models on the model. We selected three knowledge graph embedding models: TransE~\citep{bordes2013translating}, TransR~\citep{lin2015learning}, and RotatE~\citep{Sun2019Rotate}. The results on the IU-Xray dataset are 0.168/0.171/0.178 (BLEU-4) and 0.368/0.371/0.382 (CIDEr), respectively. Compared with Table~\ref{tab:ablation}, the results show that different knowledge graph embedding models affect our model's performance, but all these models outperform the model without a knowledge graph. Therefore, integrating knowledge graphs can effectively improve radiology report generation model, and higher quality knowledge graphs embeddings will lead to higher quality generated reports.

\subsection{Qualitative Results}
Figure~\ref {Fig:case} shows qualitative results of our proposed method and basic model on the MIMIC-CXR dataset. For each radiology x-ray image, our proposed method firstly retrieves top@10 related radiology records by the way described in section~\ref{sec:internal}. We only present one of the retrieved related radiology records due to figure size limit.
We randomly sample a radiology record from the MIMIC-CXR dataset as the target. The figure displays four parts of information. The first part includes the target x-ray image and reference report. The second part consists of the retrieved radiology x-ray image and report, radiology term entities extracted from the report, and corresponding specific knowledge. The rest parts include the report generated by ours and the basic model. In the figure, the same radiology diseases or terms are marked by the same colors. As shown in the figure, we highlight five types of radiology disease or terms. The retrieved specific knowledge provides related radiology relations between source radiology terms and target terms, such as \textcircled{3}\textit{effusion located at bilateral}, \textcircled{4}\textit{opacity suggestive of atelectasis}. 
Compared with the base model, the report generated by our proposed method is more accurate and covers more crucial radiology findings in the x-ray image by incorporating knowledge. The observations illustrate that our proposed method can provide accurate and reasonable radiology reports. More cases can be found in Appendix.

\section{Discussion and Conclusion}
This work proposes a novel radiology generation framework assisted by two extra knowledge, including general and specific knowledge. The general knowledge is a pre-defined knowledge graph that consists of the common knowledge for generating all radiology reports. The specific knowledge is image dependent and consists of a customized set of knowledge for current input. We also propose a  knowledge-enhanced multi-head attention mechanism to combine the structural information of knowledge and visual features. Experiments demonstrate consistent performance improvements of our proposed modules. Our model also achieves state-of-the-art performance on various metrics for both the public IU-Xray and MIMIC-CXR datasets.

However, our work is far from perfect. One major limitation of our model is that building the knowledge graph is laborious, making it hard to transfer those approaches directly to other datasets as the knowledge graph needs to be rebuilt. However, this problem also happens on other report generation models such as KERP~\citep{Li2019Knowledge}, MKG~\citep{Zhang2020when}, PPKED~\citep{Liu2021Exploring}, etc. Therefore, our next work is to develop a more generalizable model, in which we will propose a knowledge updating mechanism to learn and store medical knowledge automatically during training. It is also worth mentioning that a typical radiology report generation consists of multiple steps, and generating the findings is just a preliminary attempt to develop technology for report automation.

\section{Acknowledgments}
This work was supported by the CCF-Tencent Open Fund and the National Natural Science Foundation of China under Grant 31900979.

\bibliographystyle{model2-names.bst}\biboptions{authoryear}
\bibliography{refs}

\begin{thebibliography}{45}
\expandafter\ifx\csname natexlab\endcsname\relax\def\natexlab#1{#1}\fi
\providecommand{\url}[1]{\texttt{#1}}
\providecommand{\href}[2]{#2}
\providecommand{\path}[1]{#1}
\providecommand{\DOIprefix}{doi:}
\providecommand{\ArXivprefix}{arXiv:}
\providecommand{\URLprefix}{URL: }
\providecommand{\Pubmedprefix}{pmid:}
\providecommand{\doi}[1]{\href{http://dx.doi.org/#1}{\path{#1}}}
\providecommand{\Pubmed}[1]{\href{pmid:#1}{\path{#1}}}
\providecommand{\bibinfo}[2]{#2}
\ifx\xfnm\relax \def\xfnm[#1]{\unskip,\space#1}\fi
\bibitem[{Alfarghaly et~al.(2021)Alfarghaly, Khaled, Elkorany, Helal and
  Fahmy}]{alfarghaly2021automated}
\bibinfo{author}{Alfarghaly, O.}, \bibinfo{author}{Khaled, R.},
  \bibinfo{author}{Elkorany, A.}, \bibinfo{author}{Helal, M.},
  \bibinfo{author}{Fahmy, A.}, \bibinfo{year}{2021}.
\newblock \bibinfo{title}{Automated radiology report generation using
  conditioned transformers}.
\newblock \bibinfo{journal}{Informatics in Medicine Unlocked}
  \bibinfo{volume}{24}, \bibinfo{pages}{100557}.
\bibitem[{Alsentzer et~al.(2019)Alsentzer, Murphy, Boag, Weng, Jin, Naumann and
  McDermott}]{alsentzer2019publicly}
\bibinfo{author}{Alsentzer, E.}, \bibinfo{author}{Murphy, J.R.},
  \bibinfo{author}{Boag, W.}, \bibinfo{author}{Weng, W.H.},
  \bibinfo{author}{Jin, D.}, \bibinfo{author}{Naumann, T.},
  \bibinfo{author}{McDermott, M.}, \bibinfo{year}{2019}.
\newblock \bibinfo{title}{Publicly available clinical bert embeddings}.
\newblock \bibinfo{journal}{arXiv preprint arXiv:1904.03323} .
\bibitem[{Anderson et~al.(2018)Anderson, He, Buehler, Teney, Johnson, Gould and
  Zhang}]{anderson2018bottom}
\bibinfo{author}{Anderson, P.}, \bibinfo{author}{He, X.},
  \bibinfo{author}{Buehler, C.}, \bibinfo{author}{Teney, D.},
  \bibinfo{author}{Johnson, M.}, \bibinfo{author}{Gould, S.},
  \bibinfo{author}{Zhang, L.}, \bibinfo{year}{2018}.
\newblock \bibinfo{title}{Bottom-up and top-down attention for image captioning
  and visual question answering}, in: \bibinfo{booktitle}{Proceedings of the
  IEEE conference on computer vision and pattern recognition}, pp.
  \bibinfo{pages}{6077--6086}.
\bibitem[{Bordes et~al.(2013)Bordes, Usunier, Garcia-Duran, Weston and
  Yakhnenko}]{bordes2013translating}
\bibinfo{author}{Bordes, A.}, \bibinfo{author}{Usunier, N.},
  \bibinfo{author}{Garcia-Duran, A.}, \bibinfo{author}{Weston, J.},
  \bibinfo{author}{Yakhnenko, O.}, \bibinfo{year}{2013}.
\newblock \bibinfo{title}{Translating embeddings for modeling multi-relational
  data}.
\newblock \bibinfo{journal}{Advances in neural information processing systems}
  \bibinfo{volume}{26}.
\bibitem[{Bruno et~al.(2015)Bruno, Walker and
  Abujudeh}]{bruno2015understanding}
\bibinfo{author}{Bruno, M.A.}, \bibinfo{author}{Walker, E.A.},
  \bibinfo{author}{Abujudeh, H.H.}, \bibinfo{year}{2015}.
\newblock \bibinfo{title}{Understanding and confronting our mistakes: the
  epidemiology of error in radiology and strategies for error reduction}.
\newblock \bibinfo{journal}{Radiographics} \bibinfo{volume}{35},
  \bibinfo{pages}{1668--1676}.
\bibitem[{Chen et~al.(2020)Chen, Song, Chang and Wan}]{Chen2020Generating}
\bibinfo{author}{Chen, Z.}, \bibinfo{author}{Song, Y.}, \bibinfo{author}{Chang,
  T.H.}, \bibinfo{author}{Wan, X.}, \bibinfo{year}{2020}.
\newblock \bibinfo{title}{Generating radiology reports via memory-driven
  transformer}, in: \bibinfo{booktitle}{Proceedings of the 2020 Conference on
  Empirical Methods in Natural Language Processing (EMNLP)}, pp.
  \bibinfo{pages}{1439--1449}.
\bibitem[{Cohen et~al.(2021)Cohen, Viviano, Bertin, Morrison, Torabian,
  Guarrera, Lungren, Chaudhari, Brooks, Hashir and
  Bertrand}]{Cohen2021TorchXRayVision}
\bibinfo{author}{Cohen, J.P.}, \bibinfo{author}{Viviano, J.D.},
  \bibinfo{author}{Bertin, P.}, \bibinfo{author}{Morrison, P.},
  \bibinfo{author}{Torabian, P.}, \bibinfo{author}{Guarrera, M.},
  \bibinfo{author}{Lungren, M.P.}, \bibinfo{author}{Chaudhari, A.},
  \bibinfo{author}{Brooks, R.}, \bibinfo{author}{Hashir, M.},
  \bibinfo{author}{Bertrand, H.}, \bibinfo{year}{2021}.
\newblock \bibinfo{title}{Torchxrayvision: A library of chest x-ray datasets
  and models} \URLprefix \url{https://arxiv.org/abs/2111.00595},
  \DOIprefix\doi{10.48550/ARXIV.2111.00595}.
\bibitem[{Demner-Fushman et~al.(2016)Demner-Fushman, Kohli, Rosenman, Shooshan,
  Rodriguez, Antani, Thoma and McDonald}]{Demner-Fushman2016iu-dataset}
\bibinfo{author}{Demner-Fushman, D.}, \bibinfo{author}{Kohli, M.D.},
  \bibinfo{author}{Rosenman, M.B.}, \bibinfo{author}{Shooshan, S.E.},
  \bibinfo{author}{Rodriguez, L.}, \bibinfo{author}{Antani, S.},
  \bibinfo{author}{Thoma, G.R.}, \bibinfo{author}{McDonald, C.J.},
  \bibinfo{year}{2016}.
\newblock \bibinfo{title}{Preparing a collection of radiology examinations for
  distribution and retrieval}.
\newblock \bibinfo{journal}{Journal of the American Medical Informatics
  Association} \bibinfo{volume}{23}, \bibinfo{pages}{304--310}.
\bibitem[{Denkowski and Lavie(2014)}]{denkowski2014meteor}
\bibinfo{author}{Denkowski, M.}, \bibinfo{author}{Lavie, A.},
  \bibinfo{year}{2014}.
\newblock \bibinfo{title}{Meteor universal: Language specific translation
  evaluation for any target language}, in: \bibinfo{booktitle}{Proceedings of
  the ninth workshop on statistical machine translation}, pp.
  \bibinfo{pages}{376--380}.
\bibitem[{Gilmer et~al.(2017)Gilmer, Schoenholz, Riley, Vinyals and
  Dahl}]{gilmer2017neural}
\bibinfo{author}{Gilmer, J.}, \bibinfo{author}{Schoenholz, S.S.},
  \bibinfo{author}{Riley, P.F.}, \bibinfo{author}{Vinyals, O.},
  \bibinfo{author}{Dahl, G.E.}, \bibinfo{year}{2017}.
\newblock \bibinfo{title}{Neural message passing for quantum chemistry}, in:
  \bibinfo{booktitle}{International conference on machine learning},
  \bibinfo{organization}{PMLR}. pp. \bibinfo{pages}{1263--1272}.
\bibitem[{He et~al.(2016)He, Zhang, Ren and Sun}]{He2016deep}
\bibinfo{author}{He, K.}, \bibinfo{author}{Zhang, X.}, \bibinfo{author}{Ren,
  S.}, \bibinfo{author}{Sun, J.}, \bibinfo{year}{2016}.
\newblock \bibinfo{title}{Deep residual learning for image recognition}, in:
  \bibinfo{booktitle}{Proceedings of the IEEE conference on computer vision and
  pattern recognition}, pp. \bibinfo{pages}{770--778}.
\bibitem[{Hu et~al.(2020)Hu, Fey, Zitnik, Dong, Ren, Liu, Catasta and
  Leskovec}]{hu2020open}
\bibinfo{author}{Hu, W.}, \bibinfo{author}{Fey, M.}, \bibinfo{author}{Zitnik,
  M.}, \bibinfo{author}{Dong, Y.}, \bibinfo{author}{Ren, H.},
  \bibinfo{author}{Liu, B.}, \bibinfo{author}{Catasta, M.},
  \bibinfo{author}{Leskovec, J.}, \bibinfo{year}{2020}.
\newblock \bibinfo{title}{Open graph benchmark: Datasets for machine learning
  on graphs}.
\newblock \bibinfo{journal}{arXiv preprint arXiv:2005.00687} .
\bibitem[{Irvin et~al.(2019)Irvin, Rajpurkar, Ko, Yu, Ciurea-Ilcus, Chute,
  Marklund, Haghgoo, Ball, Shpanskaya et~al.}]{Irvin2019CheXpert}
\bibinfo{author}{Irvin, J.}, \bibinfo{author}{Rajpurkar, P.},
  \bibinfo{author}{Ko, M.}, \bibinfo{author}{Yu, Y.},
  \bibinfo{author}{Ciurea-Ilcus, S.}, \bibinfo{author}{Chute, C.},
  \bibinfo{author}{Marklund, H.}, \bibinfo{author}{Haghgoo, B.},
  \bibinfo{author}{Ball, R.}, \bibinfo{author}{Shpanskaya, K.}, et~al.,
  \bibinfo{year}{2019}.
\newblock \bibinfo{title}{Chexpert: A large chest radiograph dataset with
  uncertainty labels and expert comparison}, in:
  \bibinfo{booktitle}{Proceedings of the AAAI conference on artificial
  intelligence}, pp. \bibinfo{pages}{590--597}.
\bibitem[{Jain et~al.(2021)Jain, Agrawal, Saporta, Truong, {Nguyen Duong}, Bui,
  Chambon, Lungren, Ng, Langlotz and Rajpurkar}]{Jain2021RadGraph}
\bibinfo{author}{Jain, S.}, \bibinfo{author}{Agrawal, A.},
  \bibinfo{author}{Saporta, A.}, \bibinfo{author}{Truong, S.Q.},
  \bibinfo{author}{{Nguyen Duong}, D.}, \bibinfo{author}{Bui, T.},
  \bibinfo{author}{Chambon, P.}, \bibinfo{author}{Lungren, M.},
  \bibinfo{author}{Ng, A.}, \bibinfo{author}{Langlotz, C.},
  \bibinfo{author}{Rajpurkar, P.}, \bibinfo{year}{2021}.
\newblock \bibinfo{title}{{RadGraph: Extracting Clinical Entities and Relations
  from Radiology Reports}}.
\newblock \bibinfo{journal}{arXiv preprint arXiv:2106.14463} \URLprefix
  \url{https://doi.org/10.13026/hm87-5p47.},
  \href{http://arxiv.org/abs/arXiv:2106.14463v1}{\tt arXiv:arXiv:2106.14463v1}.
\bibitem[{Jing et~al.(2019)Jing, Wang and Xing}]{Jing2019Show}
\bibinfo{author}{Jing, B.}, \bibinfo{author}{Wang, Z.}, \bibinfo{author}{Xing,
  E.}, \bibinfo{year}{2019}.
\newblock \bibinfo{title}{Show, describe and conclude: On exploiting the
  structure information of chest x-ray reports}, in:
  \bibinfo{booktitle}{Proceedings of the 57th Annual Meeting of the Association
  for Computational Linguistics}, pp. \bibinfo{pages}{6570--6580}.
\bibitem[{Jing et~al.(2018)Jing, Xie and Xing}]{Jing2018On}
\bibinfo{author}{Jing, B.}, \bibinfo{author}{Xie, P.}, \bibinfo{author}{Xing,
  E.}, \bibinfo{year}{2018}.
\newblock \bibinfo{title}{On the automatic generation of medical imaging
  reports}, in: \bibinfo{booktitle}{Proceedings of the 56th Annual Meeting of
  the Association for Computational Linguistics (Volume 1: Long Papers)}, pp.
  \bibinfo{pages}{2577--2586}.
\bibitem[{Johnson et~al.(2019)Johnson, Pollard, Greenbaum, Lungren, Deng, Peng,
  Lu, Mark, Berkowitz and Horng}]{Johnson2019MIMIC-CXR-JPG}
\bibinfo{author}{Johnson, A.E.}, \bibinfo{author}{Pollard, T.J.},
  \bibinfo{author}{Greenbaum, N.R.}, \bibinfo{author}{Lungren, M.P.},
  \bibinfo{author}{Deng, C.y.}, \bibinfo{author}{Peng, Y.},
  \bibinfo{author}{Lu, Z.}, \bibinfo{author}{Mark, R.G.},
  \bibinfo{author}{Berkowitz, S.J.}, \bibinfo{author}{Horng, S.},
  \bibinfo{year}{2019}.
\newblock \bibinfo{title}{Mimic-cxr-jpg, a large publicly available database of
  labeled chest radiographs}.
\newblock \bibinfo{journal}{arXiv preprint arXiv:1901.07042} .
\bibitem[{Li et~al.(2018)Li, Hu, Liang and Xing}]{Li2018Hybrid}
\bibinfo{author}{Li, C.Y.}, \bibinfo{author}{Hu, Z.}, \bibinfo{author}{Liang,
  X.}, \bibinfo{author}{Xing, E.P.}, \bibinfo{year}{2018}.
\newblock \bibinfo{title}{{Hybrid retrieval-generation reinforced agent for
  medical image report generation}}, in: \bibinfo{booktitle}{Advances in Neural
  Information Processing Systems}, pp. \bibinfo{pages}{1530--1540}.
\newblock \href{http://arxiv.org/abs/1805.08298}{\tt arXiv:1805.08298}.
\bibitem[{Li et~al.(2019)Li, Liang, Hu and Xing}]{Li2019Knowledge}
\bibinfo{author}{Li, C.Y.}, \bibinfo{author}{Liang, X.}, \bibinfo{author}{Hu,
  Z.}, \bibinfo{author}{Xing, E.P.}, \bibinfo{year}{2019}.
\newblock \bibinfo{title}{Knowledge-driven encode, retrieve, paraphrase for
  medical image report generation}, in: \bibinfo{booktitle}{Proceedings of the
  AAAI Conference on Artificial Intelligence}, pp. \bibinfo{pages}{6666--6673}.
\bibitem[{Lin(2004)}]{lin2004rouge}
\bibinfo{author}{Lin, C.Y.}, \bibinfo{year}{2004}.
\newblock \bibinfo{title}{Rouge: A package for automatic evaluation of
  summaries}, in: \bibinfo{booktitle}{Text summarization branches out}, pp.
  \bibinfo{pages}{74--81}.
\bibitem[{Lin et~al.(2015)Lin, Liu, Sun, Liu and Zhu}]{lin2015learning}
\bibinfo{author}{Lin, Y.}, \bibinfo{author}{Liu, Z.}, \bibinfo{author}{Sun,
  M.}, \bibinfo{author}{Liu, Y.}, \bibinfo{author}{Zhu, X.},
  \bibinfo{year}{2015}.
\newblock \bibinfo{title}{Learning entity and relation embeddings for knowledge
  graph completion}, in: \bibinfo{booktitle}{Twenty-ninth AAAI conference on
  artificial intelligence}.
\bibitem[{Liu et~al.(2021a)Liu, Ge and Wu}]{Liu2021Competence}
\bibinfo{author}{Liu, F.}, \bibinfo{author}{Ge, S.}, \bibinfo{author}{Wu, X.},
  \bibinfo{year}{2021}a.
\newblock \bibinfo{title}{{Competence-based Multimodal Curriculum Learning for
  Medical Report Generation}}, in: \bibinfo{booktitle}{Proceedings of the 59th
  Annual Meeting of the Association for Computational Linguistics and the 11th
  International Joint Conference on Natural Language Processing (Volume 1: Long
  Papers)}, \bibinfo{publisher}{Association for Computational Linguistics},
  \bibinfo{address}{Stroudsburg, PA, USA}. pp. \bibinfo{pages}{3001--3012}.
\newblock \DOIprefix\doi{10.18653/v1/2021.acl-long.234}.
\bibitem[{Liu et~al.(2021b)Liu, Wu, Ge, Fan and Zou}]{Liu2021Exploring}
\bibinfo{author}{Liu, F.}, \bibinfo{author}{Wu, X.}, \bibinfo{author}{Ge, S.},
  \bibinfo{author}{Fan, W.}, \bibinfo{author}{Zou, Y.}, \bibinfo{year}{2021}b.
\newblock \bibinfo{title}{Exploring and distilling posterior and prior
  knowledge for radiology report generation}, in:
  \bibinfo{booktitle}{Proceedings of the IEEE/CVF Conference on Computer Vision
  and Pattern Recognition}, pp. \bibinfo{pages}{13753--13762}.
\bibitem[{Liu et~al.(2021c)Liu, Yin, Wu, Ge, Zhang and
  Sun}]{Liu2021Contrastive}
\bibinfo{author}{Liu, F.}, \bibinfo{author}{Yin, C.}, \bibinfo{author}{Wu, X.},
  \bibinfo{author}{Ge, S.}, \bibinfo{author}{Zhang, P.}, \bibinfo{author}{Sun,
  X.}, \bibinfo{year}{2021}c.
\newblock \bibinfo{title}{{Contrastive Attention for Automatic Chest X-ray
  Report Generation}}, in: \bibinfo{booktitle}{Findings of the Association for
  Computational Linguistics: ACL-IJCNLP 2021}, \bibinfo{publisher}{Association
  for Computational Linguistics}, \bibinfo{address}{Stroudsburg, PA, USA}. pp.
  \bibinfo{pages}{269--280}.
\newblock \DOIprefix\doi{10.18653/v1/2021.findings-acl.23},
  \href{http://arxiv.org/abs/2106.06965}{\tt arXiv:2106.06965}.
\bibitem[{Lu et~al.(2017)Lu, Xiong, Parikh and Socher}]{lu2017knowing}
\bibinfo{author}{Lu, J.}, \bibinfo{author}{Xiong, C.}, \bibinfo{author}{Parikh,
  D.}, \bibinfo{author}{Socher, R.}, \bibinfo{year}{2017}.
\newblock \bibinfo{title}{Knowing when to look: Adaptive attention via a visual
  sentinel for image captioning}, in: \bibinfo{booktitle}{Proceedings of the
  IEEE conference on computer vision and pattern recognition}, pp.
  \bibinfo{pages}{375--383}.
\bibitem[{Nooralahzadeh et~al.(2021)Nooralahzadeh, Gonzalez, Frauenfelder,
  Fujimoto and Krauthammer}]{Nooralahzadeh2021Progressive}
\bibinfo{author}{Nooralahzadeh, F.}, \bibinfo{author}{Gonzalez, N.P.},
  \bibinfo{author}{Frauenfelder, T.}, \bibinfo{author}{Fujimoto, K.},
  \bibinfo{author}{Krauthammer, M.}, \bibinfo{year}{2021}.
\newblock \bibinfo{title}{{Progressive Transformer-Based Generation of
  Radiology Reports}}.
\newblock \bibinfo{journal}{arXiv preprint} ,
  \bibinfo{pages}{2016--2020}\href{http://arxiv.org/abs/2102.09777}{\tt
  arXiv:2102.09777}.
\bibitem[{Papineni et~al.(2002)Papineni, Roukos, Ward and
  Zhu}]{papineni2002bleu}
\bibinfo{author}{Papineni, K.}, \bibinfo{author}{Roukos, S.},
  \bibinfo{author}{Ward, T.}, \bibinfo{author}{Zhu, W.J.},
  \bibinfo{year}{2002}.
\newblock \bibinfo{title}{Bleu: a method for automatic evaluation of machine
  translation}, in: \bibinfo{booktitle}{Proceedings of the 40th annual meeting
  of the Association for Computational Linguistics}, pp.
  \bibinfo{pages}{311--318}.
\bibitem[{Qi et~al.(2020)Qi, Zhang, Zhang, Bolton and Manning}]{qi2020stanza}
\bibinfo{author}{Qi, P.}, \bibinfo{author}{Zhang, Y.}, \bibinfo{author}{Zhang,
  Y.}, \bibinfo{author}{Bolton, J.}, \bibinfo{author}{Manning, C.D.},
  \bibinfo{year}{2020}.
\newblock \bibinfo{title}{Stanza: A {Python} natural language processing
  toolkit for many human languages}, in: \bibinfo{booktitle}{Proceedings of the
  58th Annual Meeting of the Association for Computational Linguistics: System
  Demonstrations}.
\bibitem[{Shin et~al.(2016)Shin, Roberts, Lu, Demner-Fushman, Yao and
  Summers}]{Shin2016Learning}
\bibinfo{author}{Shin, H.C.}, \bibinfo{author}{Roberts, K.},
  \bibinfo{author}{Lu, L.}, \bibinfo{author}{Demner-Fushman, D.},
  \bibinfo{author}{Yao, J.}, \bibinfo{author}{Summers, R.M.},
  \bibinfo{year}{2016}.
\newblock \bibinfo{title}{{Learning to Read Chest X-Rays: Recurrent Neural
  Cascade Model for Automated Image Annotation}}.
\newblock \bibinfo{journal}{Proceedings of the IEEE Computer Society Conference
  on Computer Vision and Pattern Recognition} \bibinfo{volume}{2016-Decem},
  \bibinfo{pages}{2497--2506}.
\newblock \DOIprefix\doi{10.1109/CVPR.2016.274},
  \href{http://arxiv.org/abs/1603.08486}{\tt arXiv:1603.08486}.
\bibitem[{Sun et~al.(2019)Sun, Deng, Nie and Tang}]{Sun2019Rotate}
\bibinfo{author}{Sun, Z.}, \bibinfo{author}{Deng, Z.H.}, \bibinfo{author}{Nie,
  J.Y.}, \bibinfo{author}{Tang, J.}, \bibinfo{year}{2019}.
\newblock \bibinfo{title}{{Rotate: Knowledge graph embedding by relational
  rotation in complex space}}.
\newblock \bibinfo{journal}{7th International Conference on Learning
  Representations, ICLR 2019} ,
  \bibinfo{pages}{1--18}\href{http://arxiv.org/abs/1902.10197}{\tt
  arXiv:1902.10197}.
\bibitem[{Vaswani et~al.(2017)Vaswani, Shazeer, Parmar, Uszkoreit, Jones,
  Gomez, Kaiser and Polosukhin}]{Vaswani2017Attention}
\bibinfo{author}{Vaswani, A.}, \bibinfo{author}{Shazeer, N.},
  \bibinfo{author}{Parmar, N.}, \bibinfo{author}{Uszkoreit, J.},
  \bibinfo{author}{Jones, L.}, \bibinfo{author}{Gomez, A.N.},
  \bibinfo{author}{Kaiser, {\L}.}, \bibinfo{author}{Polosukhin, I.},
  \bibinfo{year}{2017}.
\newblock \bibinfo{title}{Attention is all you need}, in:
  \bibinfo{booktitle}{Advances in neural information processing systems}, pp.
  \bibinfo{pages}{5998--6008}.
\bibitem[{Vedantam et~al.(2015)Vedantam, Lawrence~Zitnick and
  Parikh}]{vedantam2015cider}
\bibinfo{author}{Vedantam, R.}, \bibinfo{author}{Lawrence~Zitnick, C.},
  \bibinfo{author}{Parikh, D.}, \bibinfo{year}{2015}.
\newblock \bibinfo{title}{Cider: Consensus-based image description evaluation},
  in: \bibinfo{booktitle}{Proceedings of the IEEE conference on computer vision
  and pattern recognition}, pp. \bibinfo{pages}{4566--4575}.
\bibitem[{Vinyals et~al.(2015)Vinyals, Toshev, Bengio and
  Erhan}]{vinyals2015show}
\bibinfo{author}{Vinyals, O.}, \bibinfo{author}{Toshev, A.},
  \bibinfo{author}{Bengio, S.}, \bibinfo{author}{Erhan, D.},
  \bibinfo{year}{2015}.
\newblock \bibinfo{title}{Show and tell: A neural image caption generator}, in:
  \bibinfo{booktitle}{Proceedings of the IEEE conference on computer vision and
  pattern recognition}, pp. \bibinfo{pages}{3156--3164}.
\bibitem[{Wadden et~al.(2019)Wadden, Wennberg, Luan and
  Hajishirzi}]{wadden2019entity}
\bibinfo{author}{Wadden, D.}, \bibinfo{author}{Wennberg, U.},
  \bibinfo{author}{Luan, Y.}, \bibinfo{author}{Hajishirzi, H.},
  \bibinfo{year}{2019}.
\newblock \bibinfo{title}{Entity, relation, and event extraction with
  contextualized span representations}, in: \bibinfo{booktitle}{Proceedings of
  the 2019 Conference on Empirical Methods in Natural Language Processing and
  the 9th International Joint Conference on Natural Language Processing
  (EMNLP-IJCNLP)}, pp. \bibinfo{pages}{5784--5789}.
\bibitem[{Wang et~al.(2018)Wang, Peng, Lu, Lu and Summers}]{Wang2018TieNet}
\bibinfo{author}{Wang, X.}, \bibinfo{author}{Peng, Y.}, \bibinfo{author}{Lu,
  L.}, \bibinfo{author}{Lu, Z.}, \bibinfo{author}{Summers, R.M.},
  \bibinfo{year}{2018}.
\newblock \bibinfo{title}{Tienet: Text-image embedding network for common
  thorax disease classification and reporting in chest x-rays}, in:
  \bibinfo{booktitle}{Proceedings of the IEEE conference on computer vision and
  pattern recognition}, pp. \bibinfo{pages}{9049--9058}.
\bibitem[{Xu et~al.(2015)Xu, Ba, Kiros, Cho, Courville, Salakhudinov, Zemel and
  Bengio}]{xu2015show}
\bibinfo{author}{Xu, K.}, \bibinfo{author}{Ba, J.}, \bibinfo{author}{Kiros,
  R.}, \bibinfo{author}{Cho, K.}, \bibinfo{author}{Courville, A.},
  \bibinfo{author}{Salakhudinov, R.}, \bibinfo{author}{Zemel, R.},
  \bibinfo{author}{Bengio, Y.}, \bibinfo{year}{2015}.
\newblock \bibinfo{title}{Show, attend and tell: Neural image caption
  generation with visual attention}, in: \bibinfo{booktitle}{International
  conference on machine learning}, pp. \bibinfo{pages}{2048--2057}.
\bibitem[{Xu et~al.(2018)Xu, Hu, Leskovec and Jegelka}]{xu2018powerful}
\bibinfo{author}{Xu, K.}, \bibinfo{author}{Hu, W.}, \bibinfo{author}{Leskovec,
  J.}, \bibinfo{author}{Jegelka, S.}, \bibinfo{year}{2018}.
\newblock \bibinfo{title}{How powerful are graph neural networks?}, in:
  \bibinfo{booktitle}{International Conference on Learning Representations}.
\bibitem[{Xue et~al.(2018)Xue, Xu, Long, Xue, Antani, Thoma and
  Huang}]{Xue2018Multimodal}
\bibinfo{author}{Xue, Y.}, \bibinfo{author}{Xu, T.}, \bibinfo{author}{Long,
  L.R.}, \bibinfo{author}{Xue, Z.}, \bibinfo{author}{Antani, S.},
  \bibinfo{author}{Thoma, G.R.}, \bibinfo{author}{Huang, X.},
  \bibinfo{year}{2018}.
\newblock \bibinfo{title}{Multimodal recurrent model with attention for
  automated radiology report generation}, in: \bibinfo{booktitle}{International
  Conference on Medical Image Computing and Computer-Assisted Intervention},
  \bibinfo{organization}{Springer}. pp. \bibinfo{pages}{457--466}.
\bibitem[{Ying et~al.(2021)Ying, Cai, Luo, Zheng, Ke, He, Shen and
  Liu}]{Ying2021Do}
\bibinfo{author}{Ying, C.}, \bibinfo{author}{Cai, T.}, \bibinfo{author}{Luo,
  S.}, \bibinfo{author}{Zheng, S.}, \bibinfo{author}{Ke, G.},
  \bibinfo{author}{He, D.}, \bibinfo{author}{Shen, Y.}, \bibinfo{author}{Liu,
  T.Y.}, \bibinfo{year}{2021}.
\newblock \bibinfo{title}{Do transformers really perform bad for graph
  representation?}
\newblock \bibinfo{journal}{arXiv preprint arXiv:2106.05234} .
\bibitem[{You et~al.(2021)You, Liu, Ge, Xie, Zhang and
  Wu}]{you2021aligntransformer}
\bibinfo{author}{You, D.}, \bibinfo{author}{Liu, F.}, \bibinfo{author}{Ge, S.},
  \bibinfo{author}{Xie, X.}, \bibinfo{author}{Zhang, J.}, \bibinfo{author}{Wu,
  X.}, \bibinfo{year}{2021}.
\newblock \bibinfo{title}{Aligntransformer: Hierarchical alignment of visual
  regions and disease tags for medical report generation}, in:
  \bibinfo{booktitle}{International Conference on Medical Image Computing and
  Computer-Assisted Intervention}, \bibinfo{organization}{Springer}. pp.
  \bibinfo{pages}{72--82}.
\bibitem[{Yuan et~al.(2019)Yuan, Liao, Luo and Luo}]{Yuan2019Automatic}
\bibinfo{author}{Yuan, J.}, \bibinfo{author}{Liao, H.}, \bibinfo{author}{Luo,
  R.}, \bibinfo{author}{Luo, J.}, \bibinfo{year}{2019}.
\newblock \bibinfo{title}{Automatic radiology report generation based on
  multi-view image fusion and medical concept enrichment}, in:
  \bibinfo{booktitle}{International Conference on Medical Image Computing and
  Computer-Assisted Intervention}, \bibinfo{organization}{Springer}. pp.
  \bibinfo{pages}{721--729}.
\bibitem[{Zhang et~al.(2020)Zhang, Wang, Xu, Yu, Yuille and Xu}]{Zhang2020when}
\bibinfo{author}{Zhang, Y.}, \bibinfo{author}{Wang, X.}, \bibinfo{author}{Xu,
  Z.}, \bibinfo{author}{Yu, Q.}, \bibinfo{author}{Yuille, A.},
  \bibinfo{author}{Xu, D.}, \bibinfo{year}{2020}.
\newblock \bibinfo{title}{When radiology report generation meets knowledge
  graph}, in: \bibinfo{booktitle}{Proceedings of the AAAI Conference on
  Artificial Intelligence}, pp. \bibinfo{pages}{12910--12917}.
\bibitem[{Zhang et~al.(2021)Zhang, Zhang, Qi, Manning and
  Langlotz}]{Zhang2021Biomedical}
\bibinfo{author}{Zhang, Y.}, \bibinfo{author}{Zhang, Y.}, \bibinfo{author}{Qi,
  P.}, \bibinfo{author}{Manning, C.D.}, \bibinfo{author}{Langlotz, C.P.},
  \bibinfo{year}{2021}.
\newblock \bibinfo{title}{Biomedical and clinical english model packages for
  the stanza python nlp library}.
\newblock \bibinfo{journal}{Journal of the American Medical Informatics
  Association} \bibinfo{volume}{28}, \bibinfo{pages}{1892--1899}.
\bibitem[{Zhou et~al.(2021)Zhou, Greenspan, Davatzikos, Duncan, Van~Ginneken,
  Madabhushi, Prince, Rueckert and Summers}]{zhou2021review}
\bibinfo{author}{Zhou, S.K.}, \bibinfo{author}{Greenspan, H.},
  \bibinfo{author}{Davatzikos, C.}, \bibinfo{author}{Duncan, J.S.},
  \bibinfo{author}{Van~Ginneken, B.}, \bibinfo{author}{Madabhushi, A.},
  \bibinfo{author}{Prince, J.L.}, \bibinfo{author}{Rueckert, D.},
  \bibinfo{author}{Summers, R.M.}, \bibinfo{year}{2021}.
\newblock \bibinfo{title}{A review of deep learning in medical imaging: Imaging
  traits, technology trends, case studies with progress highlights, and future
  promises}.
\newblock \bibinfo{journal}{Proceedings of the IEEE} .
\bibitem[{Zhou et~al.(2019)Zhou, Rueckert and Fichtinger}]{zhou2019handbook}
\bibinfo{author}{Zhou, S.K.}, \bibinfo{author}{Rueckert, D.},
  \bibinfo{author}{Fichtinger, G.}, \bibinfo{year}{2019}.
\newblock \bibinfo{title}{Handbook of medical image computing and computer
  assisted intervention}.
\newblock \bibinfo{publisher}{Academic Press}.

\end{thebibliography}

\newpage
\section*{Appendix}
Each case includes a reference report written by radiologists, a report generated by our model, and a report generated by the basic model. And the bleu-4 score is provided for comparing the performance. The positive findings are highlighted in blue and the false-negative findings are highlighted in red.

\textbf{Case 1:} BLEU-4: 0.640 / 0.056 (ours / base model)\\
\textbf{Reference report:} as compared to the previous radiograph the patient has received a right-sided picc line. {\color{blue} the course of the line is unremarkable the tip of the line projects over the mid-to-low svc.} there is no evidence of complications notably no pneumothorax. unchanged appearance of the cardiac silhouette. moderate tortuosity of the thoracic aorta. {\color{red} small bilateral pleural effusions.}\\
\textbf{Generated Report by ours:} as compared to the previous radiograph the patient has received a right-sided picc line. {\color{blue} the course of the line is unremarkable the tip of the line projects over the mid svc.} there is no evidence of complications notably no pneumothorax. otherwise the radiograph is unchanged.\\
\textbf{Generated report by basic model:} ap upright and lateral views of the chest provided. there is no focal consolidation effusion or pneumothorax. the cardiomediastinal silhouette is normal. imaged osseous structures are intact. no free air below the right hemidiaphragm is seen.

\textbf{Case 2:} BLEU-4: 0.611 / 0.139 (ours / base model)\\
\textbf{Reference report:} 
midline sternotomy wires and mediastinal clips are again noted. the previously noted port-a-cath has been removed. the lungs are clear bilaterally without focal consolidation effusion or pneumothorax. cardiomediastinal silhouette is stable. {\color{blue} bony structures are intact. no free air below the right hemidiaphragm is seen.}\\
\textbf{Generated Report by ours:} 
midline sternotomy wires and mediastinal clips are again noted. there is no focal consolidation effusion or pneumothorax. the cardiomediastinal silhouette is normal. imaged osseous structures are intact. {\color{blue} no free air below the right hemidiaphragm is seen.}\\
\textbf{Generated report by basic model:} 
the patient is status post median sternotomy and cabg. the cardiac and mediastinal silhouettes are stable. no focal consolidation pleural effusion or evidence of pneumothorax is seen.

\textbf{Case 3:} BLEU-4: 0.518 / 0.383 (ours / base model)\\
\textbf{Reference report:} as compared to the previous radiograph there is no relevant change. {\color{blue} massive elevation of the left hemidiaphragm with subsequent basal areas of atelectasis.} borderline size of the cardiac silhouette. no pneumonia or other acute changes. no pleural effusions.\\
\textbf{Generated Report by ours:} as compared to the previous radiograph there is no relevant change. {\color{blue} elevation of the left hemidiaphragm with subsequent areas of atelectasis at the left lung bases.} no newly appeared parenchymal opacities. no pleural effusions. no pulmonary edema.\\
\textbf{Generated report by basic model:} 
as compared to the previous radiograph there is no relevant change. low lung volumes. moderate cardiomegaly with tortuosity of the thoracic aorta. no pleural effusions. no pulmonary edema. no pneumonia.
        
\textbf{Case 4:} BLEU-4: 0.505 / 0.400 (ours / base model)\\
\textbf{Reference report:}
as compared to the previous radiograph there is no relevant change. {\color{blue} low lung volumes} and {\color{blue} moderate cardiomegaly} without evidence of pulmonary edema or pleural effusions. {\color{red} moderate retrocardiac atelectasis.} no evidence of pneumonia.\\
\textbf{Generated Report by ours:} 
as compared to the previous radiograph there is no relevant change. {\color{blue} low lung volumes.} {\color{blue} moderate cardiomegaly} with mild fluid overload but no overt pulmonary edema. no larger pleural effusions. no pneumonia.\\
\textbf{Generated report by basic model:} 
as compared to the previous radiograph there is no relevant change. {\color{blue} moderate cardiomegaly} with signs of mild fluid overload. no pleural effusions. no pneumonia. no pneumothorax.

\textbf{Case 5:} BLEU-4: 0.473 / 0.0 (ours / base model)\\
\textbf{Reference report:}
as compared to the previous radiograph there is no relevant change. borderline size of the cardiac silhouette without evidence of pulmonary edema or pleural effusions. small pericardial fat pad on the left. known and healed left rib fracture. no evidence of pneumonia. {\color{blue} mild tortuosity of the thoracic aorta.}\\
\textbf{Generated Report by ours:} 
as compared to the previous radiograph there is no relevant change. no evidence of pneumonia. no pulmonary edema. no pleural effusions. {\color{blue} borderline size of the cardiac silhouette with tortuosity of the thoracic aorta.}\\
\textbf{Generated report by basic model:} 
in comparison with the study of there is little change and no evidence of acute cardiopulmonary disease. no pneumonia vascular congestion or pleural effusion.

\textbf{Case 6:} BLEU-4: 0.0 / 0.0 (ours / base model)\\
\textbf{Reference report:}
heart size is normal when allowances are made for prominent bilateral pericardial fat pads shown to better detail on ct abdomen of. mediastinal and hilar contours are within normal limits and without change. {\color{blue} lungs are remarkable for upper lobe predominant emphysema more severe in the right upper lobe than the left.} no new focal lung abnormalities were detected and there are no pleural effusions. mild compression deformity in the mid thoracic spine is unchanged.\\
\textbf{Generated Report by ours:} 
in comparison with study of there is little change and no evidence of acute cardiopulmonary disease. no pneumonia vascular congestion or pleural effusion.\\
\textbf{Generated report by basic model:} 
in comparison with the study of there is little change and no evidence of acute cardiopulmonary disease. no pneumonia vascular congestion or pleural effusion.

\textbf{Case 7:} BLEU-4: 0.0 / 0.088 (ours / base model)\\
\textbf{Reference report:}
prominent interstitial markings are again seen not significantly changed. there is no overt pulmonary edema. there is no pleural effusion. cardiomediastinal silhouette is stable. {\color{red} coronary artery calcifications andor stents are noted. chronic compression deformity in the lower thoracic spine.}\\
\textbf{Generated Report by ours:} 
lung volumes are low limiting assessment. allowing for this there is no focal consolidation effusion or pneumothorax. the cardiomediastinal silhouette is normal. imaged osseous structures are intact. no free air below the right hemidiaphragm is seen.\\
\textbf{Generated report by basic model:} 
lung volumes are low limiting assessment. there is no focal consolidation effusion or pneumothorax. the cardiomediastinal silhouette is normal. imaged osseous structures are intact. no free air below the right hemidiaphragm is seen.

\end{document}